\documentclass[fleqn,usenatbib]{mnras}

\usepackage{newtxtext,newtxmath}
\usepackage[T1]{fontenc}

\DeclareRobustCommand{\VAN}[3]{#2}
\let\VANthebibliography\thebibliography
\def\thebibliography{\DeclareRobustCommand{\VAN}[3]{##3}\VANthebibliography}

\usepackage{graphicx}	% Including figure files
\usepackage{amsmath}	% Advanced maths commands
\usepackage{comment}
\usepackage{siunitx}
\usepackage{newtxtext,newtxmath}
\usepackage{adjustbox}

\title[Evolving lags in Mrk~110]{
The luminosity-dependent contribution from the broad line region to the wavelength-dependent lags in Mrk~110}
\author[F. M. Vincentelli et al.]{
F. M. Vincentelli,$^{1,2}$\thanks{E-mail: federico.vincentelli@villanova.edu}
I. M$\rm^{c}$Hardy,$^{2}$, J. V.  Hernández Santisteban$^{3}$, E. M. Cackett$^{4}$, J. Gelbord$^{5}$,  \newauthor Keith Horne$^{3}$, J. A.
Miller$^{4}$,  A. Lobban$^6$
\\
$^{1}$Villanova University, Department of Physics, Villanova, PA 19085, USA\\
$^{2}$Department of Physics and Astronomy, University of Southampton, SO17 1BJ, UK\\
$^{3}$SUPA Physics and Astronomy, University of St. Andrews, North Haugh, KY16 9SS, UK\\
$^{4}$Wayne State University, Department of Physics \& Astronomy, 666 W Hancock St, Detroit, MI 48201, USA\\
$^5$Spectral Sciences Inc., 4 Fourth Avenue, Burlington, MA 01803, USA\\
$^6$European Space Agency (ESA), European Space Astronomy Centre (ESAC), 28691 Villanueva de la Cañada, Madrid, Spain
}

\date{Accepted XXX. Received YYY; in original form ZZZ}

\pubyear{2015}

\begin{document}
\label{firstpage}
\pagerange{\pageref{firstpage}--\pageref{lastpage}}
\maketitle

\begin{abstract}
{We have measured the wavelength-dependent lags between the X-ray, UV and optical bands in the 
high accretion rate ($L/L_{\rm Edd}\approx40\%$) Active Galactic Nucleus Mrk~110 during two intensive monitoring campaigns in February and September 2019.  We divide the observations into three intervals with different X-ray luminosities. The first interval, already published in Vincentelli et~al.~(2021), 
%MNRAS 504,  3, 4337)
has the lowest X-ray luminosity and did not exhibit the U-band excess positive lag, or the X-ray excess negative lag that is seen in most AGN. However, these excess lags are seen in the two subsequent intervals of higher X-ray luminosity. Although the data are limited, the excess lags appear to scale with X-ray luminosity. Our modelling shows that lags expected from reprocessing of X-rays by the accretion disc vary hardly at all with increasing luminosity. Therefore, as the U-band excess almost certainly arises from Balmer continuum emission from the broad line region (BLR), we attribute these lag changes to changes in the contribution from the BLR. The change is easily explained by the usual increase in the inner radius of the BLR with increasing ionising luminosity.
}

\end{abstract}


%%%%%%%%%%%%%%%%%%%%%%%%%%%%%%%%%%%%%%%%%%%%%%%%%%

%%%%%%%%%%%%%%%%% BODY OF PAPER %%%%%%%%%%%%%%%%%%

\section{Introduction}

  Active galactic nuclei (AGN) are believed to originate from the accretion of matter onto super-massive black holes at the center of galaxies \citep{padovani2017,eht2019_bh}. The multi-wavelength emission generated from these objects shows a complex spectral energy distribution arising from different physical processes. The two main emitting components usually invoked to explain the emission from optical to hard X-rays are a multi-color black-body spectrum arising from the accretion disc (peaking in optical and UV) and non-thermal radiation (with an X-ray power-law spectrum) from a hot compact inflow  close to the central black hole \citep{shakura1973,haardt_maraschi1991,padovani2017,noda2018}.   However,   the exact geometry of the accretion flow and its behaviour are still not fully understood.   

The strong variability of these objects can be used to constrain the structure of the accretion flow of these systems  
\citep[see e.g.][]{uttley2014}. A particularly insightful method is the study of time delays between the emission in different bands, which can been interpreted as the light travel time distance between different emitting regions \citep{blandford1982,cackett2021}. This technique, known as \lq\lq reverberation mapping\rq\rq, was initially developed to study delays between the optical-UV continuum, originating in the accretion inflow, and the emission lines \citep{peterson1993,peterson2004} from a region of orbiting and/or outflowing gas, also known as the broad-line region (BLR).   Thanks to the improvement of the photometric monitoring campaigns of AGN, reverberation mapping studies have significantly grown in the last few years, focusing on the accretion disc and opening a new way to measure the geometry of the accretion inflow \citep[][]{cackett2018,cackett2020,edelson2015,edelson2017,edelson2019,fausnaugh2016,fausnaugh2018,mchardy2014,mchardy2018,hernandez,vincentelli+2021}.

The various studies performed in the last decade have shown that even though optical/UV emission lags do broadly follow the expectation of a standard optically thick, geometrically thin accretion disc \citep[][]{shakura1973}, more physical components are required to explain the correlated variability in X-ray, UV and optical. For instance, it has been solidly established that the emission around 3600\si{\angstrom} (i.e. at the Balmer Jump) is dominated by the BLR, leading to the so called \lq\lq U-band excess\rq\rq~in almost all the observed sources \citep[][]{korista2001,korista2019,cackett2018,lawther2018}. Moreover, the X-ray vs UVW2 lags seem to show a larger-than-expected amplitude when compared to standard accretion disc models \citep[see also][]{li2021}. This led to the development of several models attempting to reproduce the lag spectrum with reprocessing driven in a lamp-post model with a relatively high corona \citep[][]{kammoun2021b}, an extended reprocessor \citep[][]{gardner-done2015}, or even with the BLR \citep[][]{korista2001,korista2019,lawther2018,cheleouce2019,netzer2020,netzer2021,cackett2021freq}.

Recent intensive monitoring observations of the high accretion rate   AGN Mrk~110 \citep[40\% Eddington ratio,][]{Meyer-Hofmeister}, showed behaviour consistent with a combination of an accretion disc and the BLR, but without an X-ray or U-band excess \citep[][]{vincentelli+2021}. In this Letter, we investigate this combination further and, in particular, the variation of the BLR contribution to the lags as a function of luminosity. We present new \textit{Swift} observations of Mrk~110, which we combine with all archival data. In Section 2 we describe the data reduction procedure. Section 3 is dedicated to the analysis of the variability and inter-band lags. Finally in Section 4 we discuss the results in terms of the currently accepted physical scenarios.

\vspace{-0.5cm}

\section{Observations.}

\subsection{Swift}
 The \textit{Neil Gehrels Swift Observatory} \citep[{\it Swift} here after][]{Gehrels2004} has observed Mrk~110 on a number of occasions
 since its launch. For our analysis we extracted all the data from the archive for both XRT and UVOT. For the XRT, we generated a time series with all the PC observations in the 0.3-10 keV range using the online XRT build-product software using the snapshot binning procedure 
 \citep[][]{evans2007}\footnote{\url{https://www.swift.ac.uk/user_objects/}}  .

 UVOT fluxes were extracted using the standard UVOT extraction software developed for UVOT lightcurves   \citep{gelbord2015}. UVOT observations can show a flux drop-off depending on the region of the detector where the source falls \citep[][]{edelson2015,hernandez}. Observations affected by this problem were flagged from their position on the detector and excluded from the analysis.

 Given the timescales involved, in order to evaluate reliable lags, intensive monitoring campaigns of at least $\approx$ one month with  almost daily sampling are required. Therefore, we selected those datasets which had at least 30 datapoints over a period of 30 days.  We identified 3 intensive monitoring time windows, which also correspond to our own proposals : the first one between MJD 58053 and 58143 \citep[already published in ][]{vincentelli+2021}, the second between  58534 and 58656 and the last between 58728  and 58797 
 (see Table~\ref{tab:obs})\footnote{These two datasets are part of the joint VLA/Swift program VLA-19A-020. Radio data will be analysed in a future work.}. We will refer to these three epochs as  E1, E2 and E3 respectively.
 
 \vspace{-0.5cm}

 \subsection{Ground Based observations}
 
During E2, we obtained quasi-simultaneous coverage from the ground, with the \textit{Las Cumbres Observatory} \citep[LCO,][]{brown2013} and the \textit{Zowada Observatory}. 

 LCO observed Mrk~110 in the SDSS \textit{griz$_s$} and Johnson \textit{BV} bands from the 1m-telescope at McDonald Observatory, USA (details of the observations are presented in Table~\ref{tab:obs}). The data were downloaded from the LCO archive which provides bias and flat-field corrected images \citep{mccully2018}. We used {\sc SExtractor} \citep{bertin1996} to perform background-subtracted aperture photometry within a 5$^{\prime\prime}$ radius centred on the source and with a global background model. Absolute flux calibration was performed using the APASS catalogue for all filters \citep{henden2018} except for $z_s$ where PANSTARRS was used \citep{Flewelling2020}.

The Zowada Observatory, a 20-inch robotic telescope in New Mexico, USA\footnote{\url{ https://clas.wayne.edu/physics/research/zowada-observatory}}, also observed Mrk~110 using SDSS griz$_s$ filters. We performed background-subtracted aperture photometry using a 5-pixel (4.4\arcsec) radius circular source aperture and annular background region with inner and outer radii of 20 and 30 pixels, respectively.  We perform relative photometry using three nearby comparison stars that are 1-3 times brighter than the target. Exposures in \textit{g}, \textit{r}, and \textit{i} were 100s, and 200s in \textit{z$_s$} and taken in groups of 5 exposures per filter.  The mean flux from exposures taken together are calculated.
 
\begin{figure}
\centering
\includegraphics[width=1.05\columnwidth]{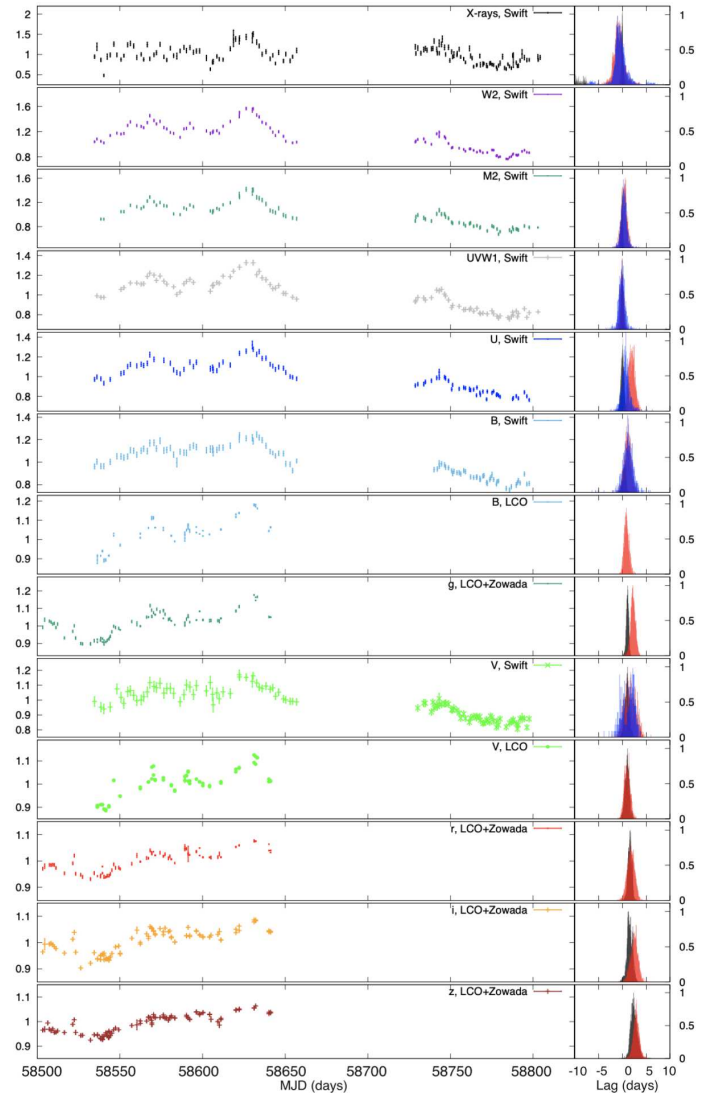}
 \caption{ \textit{Left panels:} Lightcurves from the intensive monitoring campaign of  Epoch 2 and 3. All bands are normalized to their average. \textit{Right panels:} Lag probability distribution using the  CCF FR/RSS method. {Black, red and blue histograms correspond to the lag distribution for E1, E2 and E3 respectively.}} 
\label{fig:lightcurves}
\end{figure}
\begin{table*}
		\caption{Summary of the properties of the three epochs.  {Cadence of the observations is as the average number of observations per day}. Last column reports the X-ray fractional variance \citep[see][]{vaughan2003} }

	\centering
	\begin{tabular}{lcccccccr} % four columns, alignment for each
		\hline
		Epoch  & MJD Start & MJD End &  UVOT cadence (  days$^{-1})$ &  XRT cadence (days$^{-1})$ &     Ground-based cadence (days$^{-1})$  & X-ray $F_{\rm var}$ ($\%$) \\
		\hline
		E1 & 58053 & 58143 & 2.5 & 2.8 &  1.9 & 17.6 \\
		E2 & 58534 & 58656 & 0.4 & 0.5 & 0.7  & 16.6\\
		E3 & 58728 & 58797 & 0.6 & 0.9 &  -- & 8.9\\
		\hline
    \label{tab:obs}

	\end{tabular}

\end{table*}
We inter-calibrated the LCO and Zowada griz$_s$ lightcurves using
observations separated in time by less than
$\approx$~5h. For each band we identified 8 such pairs of points and fitted a linear relation between the 
Zowada and LCO fluxes ($F_{\rm Zow}=a \, F_{\rm LCO}+b$) obtaining the slope $a$ and intercept $b$.  Using the best-fit parameters (reduced~$\chi^2$ was always close to unity), we re-scaled the Zowada fluxes to match the LCO lightcurve. The result is insensitive to inverting the order of the lightcurves, and also consistent with intercalibrations derived using \textsc{pycali} \citep[][]{li2014}. 

We note that Mrk~110 has a well known foreground nearby star \citep[$\approx$ 5.5"; see e.g. Fig 1 in ][]{bischoff1998} which is  at least 1 magnitude fainter than the AGN in all our bands. Given our aperture radius of $\approx$ 4.4-5", we expect a contribution no larger than 10\% to our target's flux. Furthermore, such a constant flux contribution will not affect the estimation of the lag.
 
\vspace{-0.5cm}

\section{Analysis}

\subsection{Lag analysis}

\begin{figure*}
	\includegraphics[width=0.9\columnwidth]{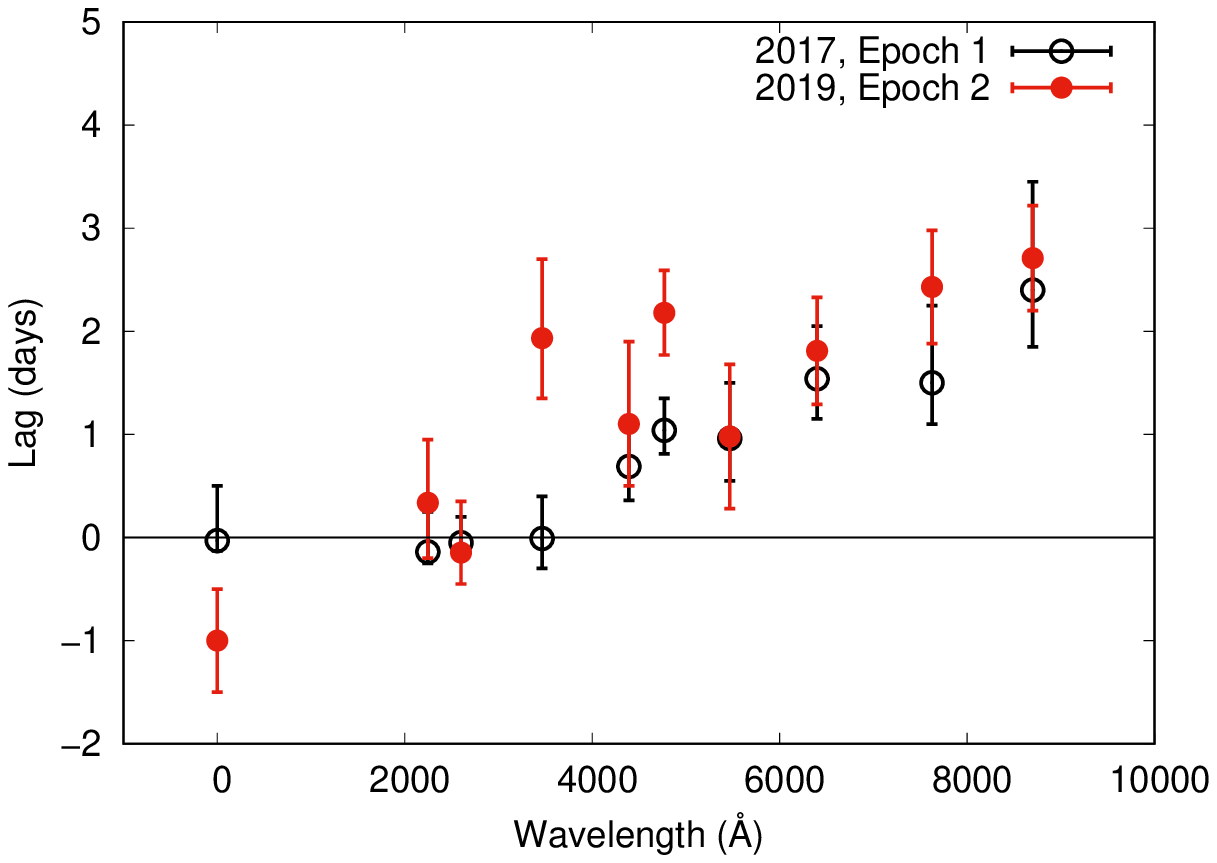}
		\includegraphics[width=0.9\columnwidth]{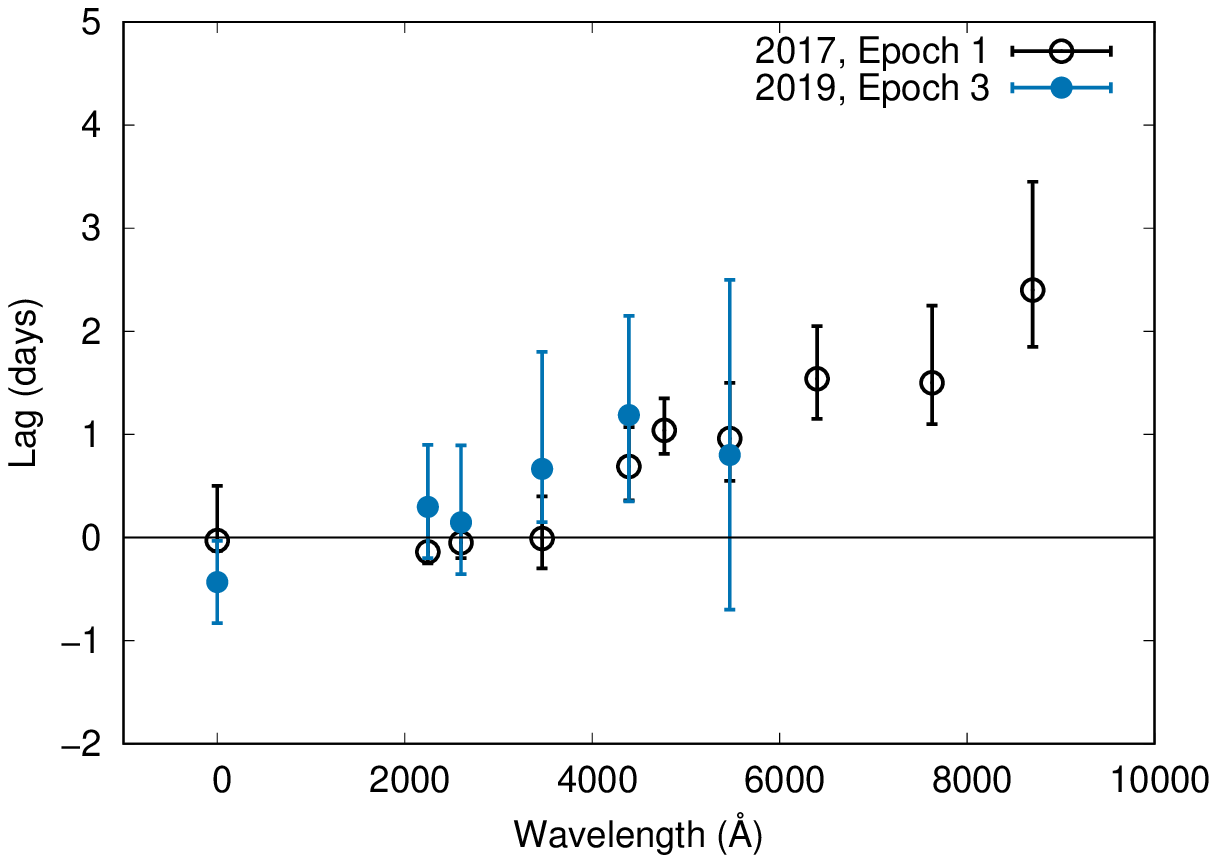}

    \caption{Lag spectrum (relative to UVW2 at 1928\AA) for Epochs 2 (red filled points, left panel) and 3 (blue filled points, right panel). While most of the bands show no significant change from Epoch 1 (empty circles in black), the X-ray and U-band lags change. }
    \label{fig:lag_spectra}
\end{figure*}

As already done for E1 \citep[][]{vincentelli+2021}, we evaluated the lags with respect to the UVW2 band  of the X-ray band and the other UV/optical filters for E2 and E3. Given the irregular sampling of the data, we applied the method known as  \lq\lq FR/RSS" to interpolate the gaps and compute the cross-correlation function (CCF) between each band and the UVW2 \citep[see e.g.][]{peterson2004}. From the randomization of the values of the interpolated data we then recovered a distribution of the interpolated CCFs centroid, which was then used to calculate the lag and its uncertainty\footnote{The calculation was done with our private \textsc{fortran} code. The software has already been tested in multiple occasions \citep[see ][]{mchardy2014,mchardy2018,pahari2020,vincentelli+2021}}. In particular for  E3 we removed the long term trend present in the data with a linear fit to the time series {\citep[see also][]{vincentelli+2021} }.  The distribution of the lag centroid for each band and the resulting lag spectra are shown in Fig.~\ref{fig:lightcurves} (right panel) and Fig.~\ref{fig:lag_spectra} respectively. The lag spectrum shows a similar shape for all the observations, apart from the  U-band, where a clear difference is seen between the three epochs, being noticeably larger in E2, where the luminosity is highest. During E2, the \textit{g} band as well shows a significant deviation from E1. 

To confirm this result,  we also computed the lags in E2 and E3 using the \textsc{javelin} software \citep[][]{zu2013}, which evaluates the lags through an MCMC simulation of top-hat impulse responses driven by a damped random walk. To avoid affecting the model, no filtering was applied in this case. The resulting lags with both methods are summarized in Table~\ref{tab:results}: these are found to be consistent within each other, confirming the evidence of an evolution of the lags with respect to the UVW2 band, in the X-ray and  U-band.

\begin{table*}
\begin{adjustbox}{angle=0}
 \begin{tabular}{ l  c | c c c c c | c c c c  r }
 % \begin{tabular}{lc|ccccc|ccccr}

\hline
%&&&& &&&& \\
&&&& \hspace{-1.5cm}Epoch 2 &&&&& Epoch 3 \\
%&&&& &&&& \\
\hline
Band & Wavelength (\si{\angstrom})  & Avg. & peak $r$ & Lag  FR/RSS   & Lag \textsc{javelin} &  & Avg. &  peak $r$ & Lag  FR/RSS   & Lag \textsc{javelin}  \\
 & &  &   & (days) &  (days)  &  &  &  & (days) &  (days)   \\
\hline
 
  X-ray & 0.25 & 1.58 & 0.50$\pm$0.06 & -1.21$\pm$0.45 &-0.91$\pm$0.10&& 1.38 & 0.43$\pm$0.06&-0.46$\pm$0.40& -0.84$\pm$0.12 \\  
 %&&&&&&&  \\
   UVW2 & 1928  & 2.54 &  -- &  --&--&& 1.87 &--&--&-- \\  
 % &&&&&&&  \\
  UVM2 & 2246  & 2.15& 0.93$\pm$0.06 & 0.34$^{+0.61}_{-0.54}$&-0.03$\pm$0.17&& 1.60  &0.81$\pm$0.06&0.30$\pm$0.55& 0.09$\pm$0.25 \\  
 % &&&&&&&  \\
   UVW1 & 2600 & 1.80 & 0.9$\pm$0.1 & -0.15$^{+0.49}_{-0.31}$&-0.27$\pm$0.25&& 1.40 &0.83$\pm$0.06 &-0.15$\pm$0.59 & -0.28 $\pm$0.32 \\
% &&&&&&&  \\
  U & 3465  & 1.26& 0.91$\pm$0.06  & 1.93$^{+0.77}_{-0.58}$ & 1.49$\pm$0.45 && 0.99 &0.57$\pm$0.11& 0.67$_{-0.72}^{+0.68}$& 1.11$\pm$0.43\\ 
%   &&&&&&&  \\

   B & 4392   &  0.71 & 0.86 $\pm$0.05 & 1.10$^{+0.79}_{-0.60}$  &1.09$\pm$0.42&& 0.57 &0.63$\pm$     0.10& 1.18$_{-0.88}^{+0.92}$& 1.81$\pm$0.58 \\ 
%      &&&&&&&  \\
  V & 5468   & 0.53 & 0.73$\pm$0.11 &1.61 $\pm$1.25 &0.70$\pm$0.72&& 0.45 & 0.6$\pm$0.1 & 0.87$_{-1.56}^{+1.84}$& 1.87$\pm$1.01  \\
%  &&&&&&&  \\

  \hline
   B & 4392 & 0.64 &0.91$\pm$0.02& 0.85$\pm$0.53 &0.3$\pm$0.2&&&&&  \\
%   &&&&&&&  \\
 g & 4770 & 0.55 & 0.89$\pm$0.03& 2.18$\pm$0.41&1.94$_{-0.26}^{+0.44}$&&&&&  \\
%   &&&&&&&  \\

 V & 5468 & 0.52 &0.88$\pm$0.03& 0.98$\pm$0.60 &0.81$\pm$0.05&&&&&  \\
 %  &&&&&&&  \\
 r &  6400  & 0.52 &0.80$\pm$0.04&1.81$\pm$0.59&1.83$\pm$0.13&&&&&  \\
%   &&&&&&&  \\
 i & 7625 & 0.32 &0.84$\pm$0.06& 2.51$\pm$0.55 &2.41$\pm$0.08&&&&&  \\
%   &&&&&&&  \\
 z & 8700 & 0.24 & 0.81 $\pm$0.08  &2.86$\pm$0.51 &2.96$\pm$0.10&&&&&  \\
 \hline
\end{tabular}
 \end{adjustbox}
         \caption{Results of the lag  with respect to the UVW2 band computed between different bands. After band name and wavelength we report the average value of the time series (in counts~s$^{-1}$ for X-rays and in $10^{-14}$ erg cm$^{-2}$ s$^{-1}$ \si{\angstrom}$^{-1}$ for the UV and optical bands), the peak correlation coefficient, the lag computed with the FR/RSS method and the lag evaluated through  \textsc{javelin}. }
    \label{tab:results}
\end{table*}{}

\vspace{-0.5cm}

\subsection{X-ray spectral analysis}

 We also analysed the time average  X-ray spectra for the three epochs. Following \citet[][]{vincentelli+2021} and \citet[][]{porquet2021} we used a phenomenological model, fitting the data with Comptonised emission plus a black-body component for the soft excess. Moreover we also added a Gaussian component to reproduce the iron line contribution at $\approx$6.4 keV. The overall model was  \textsc{zphabs} $\times$ [\textsc{pexrav+zgauss+zbbody}]. The spectra do not display any significant absorption throughout the three epochs, but show a clear change in flux which is modelled fairly straightforwardly just by changes in the normalisations of the power-law and soft-excess (see Table~\ref{tab:xray_fit}). Thus the change in flux which we observe in each of the three epochs and which we plot in Fig.~\ref{fig:lag_vs_flux} represents a change in the ionising luminosity of the source.

\begin{figure} 
\centering
\includegraphics[width=0.95\columnwidth]{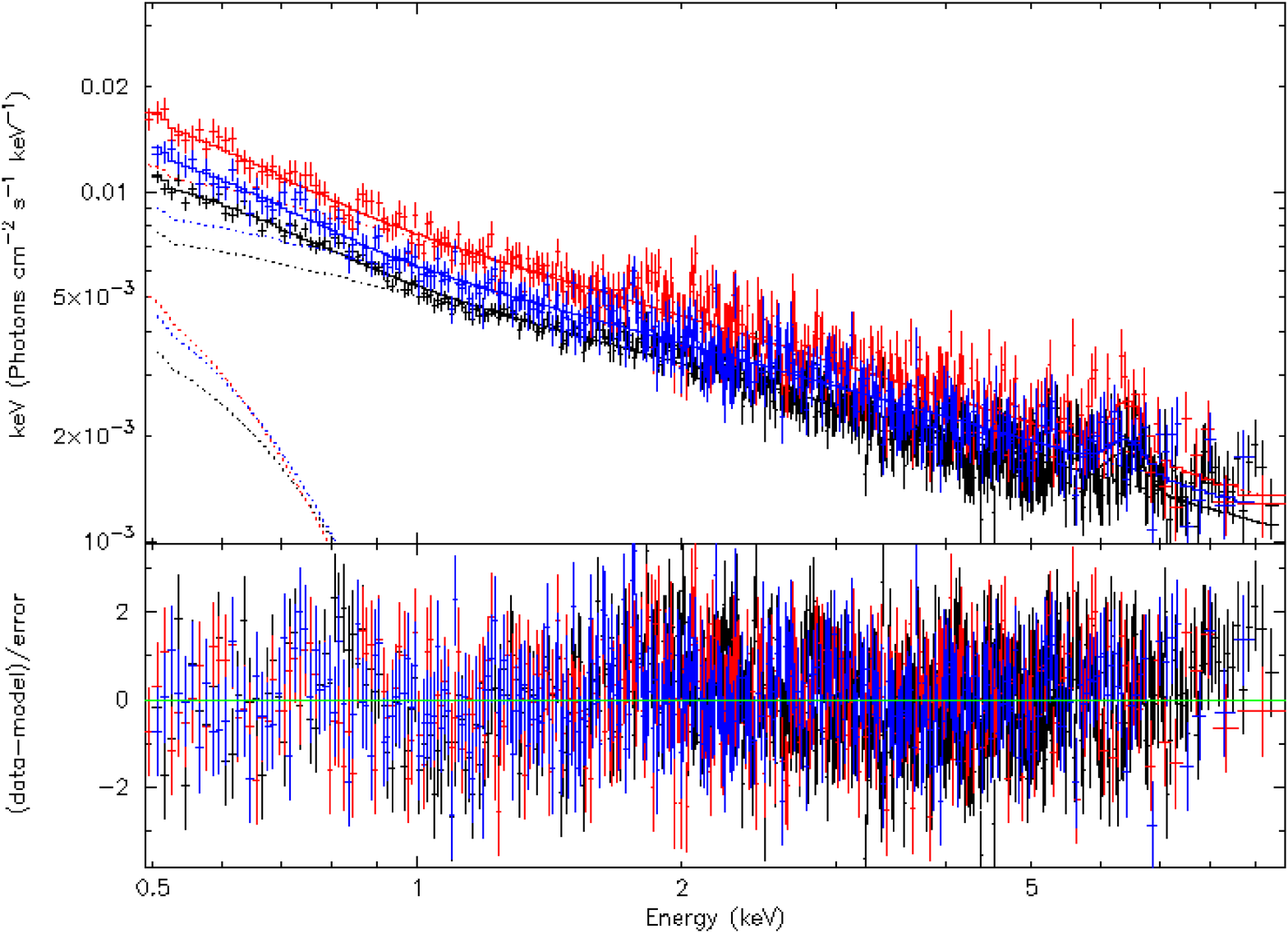}
 \caption{ Average 0.5-10~keV Swift XRT spectra for the three epochs. The lack of absorption shows that the change in flux is intrinsic to the source. {Black, red and blue spectra correspond to E1, E2 and E3, respectively}.} 
\label{fig:lc_short}
\end{figure}

\begin{table*}
	\centering

	\begin{tabular}{llccc} % four columns, alignment for each
		\hline
		Component &Parameter (units)  & E1 & E2 &  E3  \\
				\hline
		\textsc{zbbody} %\\
		& $kT$ (keV)& 0.11$\pm$ 0.05 & $0.11\pm0.01$ & 0.11 $\pm$ 0.01 \\
		& Norm. ($10^{-5}$) & 5.9 $\pm$ 0.4 & 8.7 $\pm$ 1.0 & 7.4 $\pm$ 1.0  \\
\hline
		\textsc{Pexrav} %\\
		& $\Gamma$ & 1.66$\pm$ 0.02 & $1.75\pm0.03$ & 1.68 $\pm$ 0.03 \\
		& Norm. ($10^{-3}$) & 5.2 $\pm$ 0.1 & 7.6 $\pm$ 0.2 & 6.0 $\pm$ 0.1  \\
 \hline
		\textsc{zgauss} %\\

& Energy (keV) & 6.6$\pm$ 0.1 & -- & --	\\
& $\sigma$ (keV) & 0.3$\pm$ 0.2 & -- & --\\	
& Norm. ($\times 10^{-5}$) & 4.0 $\pm$ 0.2 & 8.3 $\pm$ 1.0 & 3.8 $\pm$ 3.0  \\
		\hline
$\chi^2/{\rm dof}$  &  $1.05=(1322/1254)$ \\
		\hline
Flux ($0.5-2$~~~keV) & ($10^{-11}$~erg~cm$^{-2}$~s$^{-1}$) &
 1.7 & 2.4 & 1.9 \\
Flux (~~~$2-10$~keV) &
($10^{-11}$~erg~cm$^{-2}$~s$^{-1}$) &
 2.2 & 2.8 & 2.5 \\
Flux ($0.5-10$~keV) & ($10^{-11}$~erg~cm$^{-2}$~s$^{-1}$) &  
 3.9 & 5.2 & 4.4 \\
 
		\hline

 	\end{tabular}
 		\caption{Summary of the best-fit parameters of the XRT spectra for epochs E1, E2 and E3. Data were fitted with the mode: \textsc{zphabs} $\times$ [\textsc{pexrav+zgauss+zbbody}].  Due to the statistics of the data, following \citet[][]{vincentelli+2021} and \citet[][]{porquet2021}, we fixed the following paremeters: $N_{\rm H}= 1.27\times 10^{20}{\rm cm}^{-2}$; $z=0.035$; $E_{\rm cut}=120$~keV; ref.~fract. = 0.2; He~abund (elements heavier than He) = 1; Fe~abund=1. } 
	\label{tab:xray_fit}
\end{table*}

\begin{figure} 
\centering
\includegraphics[width=1.\columnwidth]{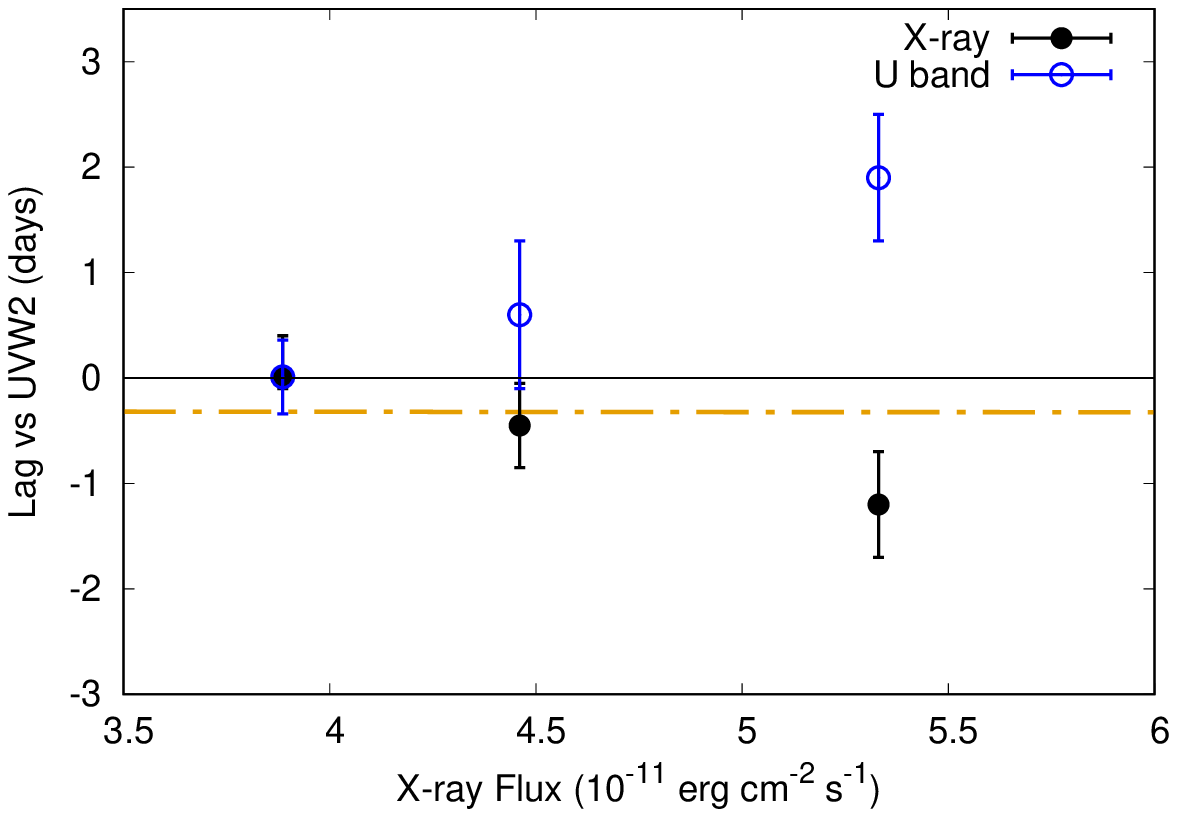}
 \caption{ X-ray vs UVW2 band lag (black) and and U  vs UVW2 band lag (blue) versus the average X-ray flux (0.5-10~keV) during the three monitoring epochs. Both lags increase their amplitude with the X-ray flux, suggesting a connection between the two bands.  The dashed-dotted line shows the X-ray lag versus X-ray flux relation predicted by the disc reprocessing model developed by \citet[][]{kammoun2021b,kammoun2021} using Mrk~110 physical parameters with a fixed coronal height of $10\,R_{\rm G}$. In this model the predicted lag variations are negligible. } 
\label{fig:lag_vs_flux}
\end{figure}

\vspace{-0.5cm}

\subsection{Lags vs Flux}

In Fig. \ref{fig:lag_vs_flux} we plot the U-band and the X-ray lag as a function of the X-ray flux. 
The (positive) U-band lag increases and the (negative) X-ray lag decreases with X-ray flux. Interestingly we also notice that the peak correlation coefficient between X-ray and UVW2 increases with X-ray flux. We also checked but found no similar relations for other quantities such as the UVW2 flux or 
the X-ray fractional variance $F_{\rm var}$ \citep[see][]{vaughan2003}, listed in Table~\ref{tab:obs}.
 To quantify the evidence for lag variations, we fit the 3 lag measurements with a constant. For the U-band the constant lag model is rejected with high confidence ($\chi^2/{\rm dof} \simeq4.1$, $p\simeq0.017$). 
 The evidence for variations 
 in the X-ray lag is weaker ($\chi^2/{\rm dof}=2.69$, $p\simeq0.13$).

\vspace{-0.5cm}

\section{Discussion and Conclusions}

We have measured wavelength-dependent lags, from the X-rays through to the \textit{z$_s$}-band, in the narrow line Seyfert I galaxy Mrk110 on three separate epochs, each at different average flux levels. In the first epoch E1 \citep[][]{vincentelli+2021}, which had the lowest flux level, we did not detect any U-band excess lag, such as is usually attributed to reprocessed light from the BLR \citep[e.g.][]{korista2001,korista2019,cackett2018,edelson2019}.  We also did not find any excess lag between the X-rays and UVW2 band which is, again, commonly found in AGN. 
{This lag spectrum was fully consistent with standard accretion disc predictions.}
However in epochs E2 and E3, where the average flux level was higher (highest in E2), excess lags were seen in both the U-band and X-ray band. Moreover, we find that both  lags increase with increasing X-ray luminosity (Fig.\ref{fig:lag_vs_flux}).  

Incidentally, we note here that we measure all lags relative to UVW2. The reason is that most previous work  \cite[e.g. see summary in Figures 16 and 17 of][]{mchardy2018}, shows that the lag spectrum in the UV and optical bands is relatively smooth, with no large discontinuities and, apart from scaling with mass and accretion rate, is similar in most AGN, indicating a similar origin for the light in those bands. However there is often a discontinuity in this lag spectrum when extended to the X-ray bands, and that discontinuity differs from one AGN to another. Thus there may be differences in structure between AGN which leads to different X-ray to UVW2 lags. 

Considering first the U-band excess lag, our disc modelling, which we have used in a number of previous papers, shows that the ionising luminosity is not a critical parameter in determining lags from accretion discs.  For example, increasing the ionising luminosity by a factor 3 increases the model lags by less than 1 per cent \citep[see e.g.][caption to Table~3]{mchardy2018}. A very similar result is shown in \citet[][see also the X-ray vs UVW2 lag prediction in Fig. \ref{fig:lag_vs_flux}]{kammoun2021b}.  Although no U-band excess was seen in the E1 observations, the optical lightcurves showed a very long response to a large X-ray flare \citep[][]{vincentelli+2021}, only explainable by reprocessing in the BLR. Thus the increase in U-band lag with increasing luminosity is very likely to be due to an increase in the BLR inner radius with increasing ionising luminosity \citep[e.g.][]{peterson2002,bentz2013,ilic2017}.

In principle, with enough measurements of lags at different luminosities, it should be possible to disentangle the separate contributions to the lags as the disc component is essentially constant and the BLR contribution will vary as  $\approx L_{\rm ion}^{1/2}$. However the number of epochs is not sufficient to perform a more detailed analysis.

We next consider the observation that the X-rays come before the UVW2 by an amount that depends on the X-ray luminosity  (Fig.~\ref{fig:lag_vs_flux}). Several explanations are possible. \citet{gardner2017}, assuming that all of the reprocessed light comes from the disc, suggest that the extra X-ray to UV lag is caused by absorption and scattering of the X-rays by the inner edge of the accretion disc. If so, then an increase in X-ray luminosity from the central source might lead to increased inflation of the inner disc, and more absorption/scattering, resulting in an increased lag. It is also possible to increase the lag between the X-rays and all of the reprocessed emission simply by moving the X-ray source further away. If we again assume that the reprocessed light all comes from the disc, then increasing the X-ray coronal size to move the X-ray centroid further from the disc, could provide an answer. According to Eqn.~8 in \citet[][]{kammoun2021}, in order to reproduce the X-ray-UVW2 lag of $\approx$ 1 day the height of the corona needs to be $>60\,R_{\rm G}$.  However, coronal height measurements from X-ray reverberation mapping typically show heights not larger than $20\,R_{\rm G}$ \citep[see e.g.][]{Emmanoulopoulos2014,alston2020}.  On the other hand,  X-ray spectral analysis of Mrk~110 \citep[][]{porquet2021}, are better fitted at higher fluxes by a larger corona (i.e. from $\approx20$  to $\approx40\,R_{\rm G}$)

It has also been proposed that the UV/optical lags may arise entirely from the BLR \citep[see e.g.][]{cheleouce2019,netzer2021} or at least be dominated by it \citep[][]{cackett2021freq}.  In this case, therefore,  the light travel time from the inner accretion flow to the BLR inner radius would correspond to the X-ray to UVW2 lag and not the UVW2 to U-band.  The change in the U-band lag would instead be due to an increasing contribution of the diffuse continiuum (DC) at the Balmer jump. 

We note that, following a drop after the g-band, the lags increase at longer wavelengths, with the lag at \textit{i} and \textit{z$_s$}  bands being comparable to that in the observed in the U-band. {Even though no significant variation is observed between E2 and E1 (which is most probably \lq\lq disc-dominated")}, we note that such a behaviour is expected also from the DC from the BLR \citep[see e.g.][]{korista2019}. Therefore  it is possible that all of the lags come from reprocessing in the BLR. We remind readers that this deduction does not imply no contribution from the disc, as the lags are not additive. If there is a smaller contribution from the disc, we will still measure the larger one from the BLR. The relative importance of the two components depends more on the relative flux received from each one. Further simulations, beyond the aim of this paper, are required in order to quantitatively disentangle the contribution between these two components.

 {This result shows for the first time that the BLR contribution can vary in a single object, confirming the importance of considering the effect of emitting components different from the disc when studying the lag phenomenology of these objects. } 
In particular we indicate two possible explanations for the observed behaviour: an evolving accretion flow with accretion rate, pushing away the inner radius of the BLR, or the BLR being responsible for the whole lag spectrum. New multi-epoch observations are required to disentangle the dependence of the lags on flux and timescales, and reach a  more detailed understanding of their physical origin.

\section*{Acknowledgements}
{The authors thank the anonymous referee for the valuable comments which improved the quality and clarity of the manuscript.} FMV and IMH were supported for this work by the STFC  grant ST/R000638/1. JVHS and KDH acknowledges support from STFC grant ST/R000824/1. EMC \& JM gratefully acknowledge support from the NSF through grant AST-1909199.

This work makes use of observations from the Las Cumbres Observatory global telescope network.

%%%%%%%%%%%%%%%%%%%%%%%%%%%%%%%%%%%%%%%%%%%%%%%%%%
\section*{Data Availability}

Raw Swift data is available on the \textsc{heasarc} website \url{https://heasarc.gsfc.nasa.gov/docs/archive.html}.
LCO data is available through its data archive at \url{https://archive.lco.global/}. UVOT and intercalibrated LCO+Zowada data can be downloaded from Zenodo \url{https://doi.org/10.5281/zenodo.5793147}.
%The inclusion of a Data Availability Statement is a requirement for articles published in MNRAS. Data Availability Statements provide a standardised format for readers to understand the availability of data underlying the research results described in the article. The statement may refer to original data generated in the course of the study or to third-party data analysed in the article. The statement should describe and provide means of access, where possible, by linking to the data or providing the required accession numbers for the relevant databases or DOIs.



\bibliographystyle{mnras}
\bibliography{bib} % if your bibtex file is called example.bib

%\begin{thebibliography}{99}
%\end{thebibliography}

%%%%%%%%%%%%%%%%%%%%%%%%%%%%%%%%%%%%%%%%%%%%%%%%%%

%%%%%%%%%%%%%%%%% APPENDICES %%%%%%%%%%%%%%%%%%%%%

%%%%%%%%%%%%%%%%%%%%%%%%%%%%%%%%%%%%%%%%%%%%%%%%%%

% Don't change these lines
\bsp	% typesetting comment
\label{lastpage}
\end{document}